\documentclass[%
 reprint,
superscriptaddress,
 amsmath,amssymb,
 aps,
]{revtex4-2}

\usepackage{graphicx}
\usepackage{dcolumn}
\usepackage{bm}

\usepackage{amsmath}
\usepackage{mathtools}
\usepackage{amsmath,amssymb}
\usepackage{physics}
\usepackage{graphicx}
\usepackage{booktabs}
\usepackage{mathrsfs}
\usepackage{dsfont}
\usepackage{hyperref}
\hypersetup{
     colorlinks=true,
     linkcolor=blue,
     filecolor=blue,
     citecolor = blue,      
     urlcolor=blue,
     }

\begin{document}

\preprint{APS/123-QED}

\title{Device-independent quantum key distribution based on Bell inequalities with more than two inputs and two outputs}

\author{Junior R. Gonzales-Ureta}
\email{junior.gonzales@fysik.su.se}
\affiliation{%
 Department of Physics, Stockholm University, 10691 Stockholm, Sweden
}

\author{Ana Predojevi\'{c}}
\email{ana.predojevic@fysik.su.se}
\affiliation{%
 Department of Physics, Stockholm University, 10691 Stockholm, Sweden
}%

\author{Ad\'{a}n Cabello}
\email{adan@us.es}
\affiliation{%
 Departamento de F\'{\i}sica Aplicada II, Universidad de Sevilla, E-41012 Sevilla, Spain
}%
\affiliation{Instituto Carlos I de F\'{\i}sica Te\'{o}rica y Computacional, Universidad de Sevilla, E-41012 Sevilla, Spain}



\date{\today}
             

\begin{abstract}
Device-independent quantum key distribution (DI-QKD) offers the strongest form of security against eavesdroppers bounded by the laws of quantum mechanics. However, a practical implementation is still pending due to the requirement of combinations of visibility and detection efficiency that are beyond those possible with current technology. This mismatch motivates the search for DI-QKD protocols that can close the gap between theoretical and practical security. In this work, we present two DI-QKD protocols whose security relies on Bell inequalities  with more than two inputs and two outputs. We show that, for maximally entangled states and perfect visibility, a protocol based on a Bell inequality with three inputs and four outputs requires a slightly lower detection efficiency than the protocols based on Bell inequalities with two inputs and two outputs. 
\end{abstract}

\maketitle



\section{Introduction}


\subsection{Context}


In the device independent quantum key distribution (DI-QKD) paradigm \cite{ekert1991quantum,mayers1998quantum,mayers2003self,barrett2005no,pironio2009device}, two parties (Alice and Bob), using only the input-output correlations obtained in a Bell inequality-like experiment, aim to generate a cryptographic key while quantifying the amount of information available to an eavesdropper (Eve) bounded by the laws of quantum mechanics.

Suppose that each party can choose between $n$ different measurements and each measurement yields one of $m$ possible outcomes.  Then, there are $n^2$ probability distributions each of them specified by $m^2$ probabilities, which can be arranged in a list ${\textbf{P}} =\{p(a,b|x,y)\}$, where $p(a,b|x,y)$ is the probability that Alice obtains outcome $a$ for measurement $x$ and Bob obtains outcome $b$ for measurement $y$. ${\textbf{P}}$ is usually referred to as a behavior \cite{cirelson1993some} and is the tool used by Alice and Bob to check the security of the key. 

Since, a behavior does not refer to any particular physical realization or quantum system of a given dimension, in DI-QKD Alice and Bob can treat their preparation and measurement devices as black boxes in which the measurement choices are the inputs and the measurement outcomes are the outputs. A quantum realization $\textbf{Q}$ of a behavior ${\textbf{P}}$ consists of a quantum state $\rho$, a set of measurements for Alice $\{M_{a|x}\}$, and a set of measurements for Bob $\{M_{b|y}\}$.

DI-QKD offers security under a minimal set of assumptions \cite{zap2019long,murta2019towards} and against a wide range of side channel attacks \cite{pironio2009device}. Side channel attacks have been shown to compromise the security of some commercial implementations of QKD \cite{fung2007phase,lydersen2010hacking,zhao2008quantum}.

However, DI-QKD still has to overcome several challenges before being technologically feasible. In what follows, we focus on photonic implementations, given that photons are the most suitable physical systems to carry out QKD in real life. 

One of the most significant obstacles is that DI-QKD requires combinations of overall detection efficiency $\eta$ and visibility $V$, which are very difficult to achieve with current technology. 
The overall detection efficiency $\eta$ is defined as the probability of detecting a photon emitted by the source. The visibility $V$ is defined assuming that the targeted state $\ket{\psi}$ is affected by white noise. That is, assuming that the actual state is of the form
$\rho = V \ket{\psi}\bra{\psi} + \frac{1-V}{d^2} \mathds{1}$, 
where $d$ is the dimension of the local systems.

For example, for the most studied DI-QKD protocols, which are those based on the Clauser-Horne-Shimony-Holt (CHSH) \cite{CHSH1969} Bell inequality, the minimum detection efficiency required to distill a secret key was initially found to be $\eta = 0.924$, assuming $V=1$ \cite{pironio2009device}.
Subsequently, it was shown that, using partially entangled states, the threshold can be reduced to $\eta =  0.865$, again with $V=1$ \cite{woodhead2020device,tan2019computing,brown2020computing}.

Recently, various efforts have been made to reduce the detection efficiency threshold \cite{woodhead2020device,sekatski2020device,schwonnek2020robust,ho2020noisy}. The most successful one \cite{woodhead2020device} reported detection efficiency thresholds of $\eta = 0.8257$ for $V=1$ and $\eta = 0.8757$ for $V=0.99$. However, any of these requirements is still very difficult to meet. 

To date, the best combination of parameters ($\eta,V$) reported in photonic experiments are ($0.763, \approx0.99$) \cite{shalm2021device}, ($0.774,0.99$) \cite{giustina2015}, and ($0.8411,0.9875$) \cite{liu2021device}. 

All the examples just mentioned refer to CHSH inequality-based DI-QKD protocols. This Bell inequality has two inputs and two outputs per party. 
There, Jordan's lemma (or similar arguments) \cite{jordan1875essai,tsirelson1993some,masanes2006asymptotic,pironio2009device} offers a convenient reduction of the problem to a two-qubit system. This reduction allowed Pironio {\em et al.} \cite{pironio2009device} to derive an analytical tight bound on the quantum conditional entropy Alice-Eve $H(A|E)$ as a function of the value of the Bell parameter $S = \langle A_1 B_1 \rangle + \langle A_1 B_2 \rangle + \langle A_2 B_1 \rangle - \langle A_2 B_2 \rangle$ of the CHSH inequality 
\begin{equation}
\label{eq01}
    H(A|E) \geq 1- h\left(\frac{1+\sqrt{S^2/4-1}}{2}\right).
\end{equation}

The quantum conditional entropy quantifies the strength of the correlations between Alice and Eve, and hence the secrecy of the key. Analytical bounds for the quantum entropy like the one in Eq.~\eqref{eq01} are only known for a few cases \cite{woodhead2020device,ho2020noisy}. In addition to Jordan's Lemma, some previous results were crucial to derive these analytical bounds. One of them is the possibility to calculate the maximum violation of a Bell inequality that a two-qubit state can attain \cite{horodecki1995violating,acin2012randomness}. 

\subsection{Approach}

The motivation for the present work is the observation that the minimum detection efficiency for reaching  detection-loophole-free Bell tests can decrease when Bell inequalities with more inputs and outputs are considered \cite{massar2002nonlocality,massar2002bell}.
For example, Massar  \cite{massar2002nonlocality} proved that entangled states of large local dimension $d$ and $2^d$ measurements per party require lower detection efficiencies. Specifically, in that case, the detection efficiency threshold decreases exponentially with the dimension of the state as $\eta = d^{1/2} 2^{-0.0035 d}$. This means that loophole-free Bell tests with arbitrary low detection efficiency are, in principle, possible.

In addition, it has also been shown that high dimensional entangled states are more robust against noise than two-qubit systems \cite{collins2002bell,kaszlikowski2000violations,chen2001entangled}. Correspondingly, violations of Bell inequalities with lower visibilities are achievable. Furthermore, QKD protocols have been recently shown that benefit from the resistance to noise exhibited by high dimensional systems \cite{mirdit2021}. 

Therefore, a natural question is whether similar benefits may occur in DI-QKD protocols based on Bell inequalities with more than two inputs and two outputs. In this article, we present to two DI-QKD protocols of this type and explore their performance using the technique developed by Brown {\em et al.}~\cite{brown2020computing} to estimate \( H(A|E)\) in a device-independent way (for details, see Appendix~\ref{app.1}). This technique, together with the one in \cite{tan2019computing} have shown significant improvements with respect to the min-entropy approach \cite{masanes2011secure,konig2009operational}. 



\section{The two DI-QKD protocols}


\subsection{General considerations}

Here, we introduce two DI-QKD protocols. Each of them is constructed around a Bell inequality that has some specific features that make it worth consideration for DI-QKD. 

Both protocols follow the structure of the protocol by Pironio {\em et al.} \cite{pironio2009device}. In each measurement round, Alice randomly chooses one of \(n\) inputs \(x \in \{1,\dots, n\}\), while Bob randomly chooses one of $n+1$ inputs \(y \in \{ 1,\dots, n+1\}\). Each measurement has $m$ possible outputs  \(a, b \in \{ 1,\dots,m\}\). The raw key is obtained from the rounds where \(x=1\) and \(y=n+1\), while the other rounds are used to characterize the behavior. 

Our first protocol has $n=4$ and $m=2$, while the second one has $n=3$ and $m=4$. In both cases, to generate the behaviors {\bf P}, we use quantum realizations {\bf Q} with entangled states of two ququarts (i.e., quantum systems of dimension four). 
Using a particular realization ensures that the statistics obtained corresponds to a valid quantum probability distribution. 

The information revealed by the behavior allows Alice and Bob to bound $H(A|E)$, which quantifies the information available to an eavesdropper. If the behavior exhibits sufficiently strong correlations between Alice and Bob and sufficiently weekly correlations between each of them and an eavesdropper, then the raw key can be turned into a shared secret key by using classical error correction and privacy amplification \cite{pironio2009device,woodhead2020device}.

Unlike, Eq.~(\ref{eq01}), where $H(A|E)$ is bounded solely by the value of the Bell parameter, in our case, we use the complete behaviour to estimate $H(A|E)$. The behaviour contains much more information than the value of the Bell parameter, thus allowing a more accurate evaluation of $H(A|E)$ \cite{Nieto_Silleras_2014}.


The asymptotic secret key rate against collective attacks is given by the Devetak-Winter formula \cite{devetak2005distillation}
\begin{equation}
    r_{\text{DW}} \geq H(A_1|E) - H(A_1|B_{n+1}),
\end{equation}
where $H(A_1|B_{n+1})$ is the conditional Shannon entropy, which quantifies the strength of the correlations between Alice and Bob. $H(A_1|B_{n+1})$ is calculated from the behavior as follows:
\begin{equation}\label{eq.hab}
\begin{split}
    H(A_1|B_{n+1}) = &-\sum_{a,b} p(a,b|1,n+1) \log_2 p(a,b|1,n+1)\\
    &+\sum_{b} p_{B}(b|n+1) \log_2 p_{B}(b|n+1),
\end{split}
\end{equation}
where $p_{B}(b|n+1)$ is the probability that Bob obtains outcome $b$ when the input is $n+1$.

We investigate how the secret key rate changes when white noise and limited detection efficiency are taken into account. 

To study the effect of the limited detection efficiency, we assume that Alice and Bob have the same detection efficiency $\eta_a = \eta_b = \eta$ for all detectors and measurement settings. We also assume that Alice and Bob map the nondetection events, denoted $\perp$, to the \(m^{th}\) output. In this way, the violation of the Bell inequality by itself assures that there is no local hidden variable model even when the detectors have limited detection efficiency  \cite{brunner2014bell,ekert2014ultimate}. The limited detection efficiency transforms the probabilities as follows:
\begin{equation}
    \begin{split}
        p^{\prime}(a,b|x,y) \xrightarrow[]{} & ~\eta^2 p(a,b|x,y) + \eta \Bar{\eta} [ \delta_{a,m} p_{B}(b|y) +\\ &\delta_{b,m} p_{A}(a|x)] + \delta_{a,m} \delta_{b,m}\Bar{\eta}^2,
    \end{split}
\end{equation}
where \(\Bar{\eta} = 1-\eta\), and $\delta_{x,y}$ is the Kronecker delta function.


\subsection{Protocol based on \(I^4_{4422}\) using maximal ququart-ququart entanglement}


The first protocol is constructed around the $I^4_{4422}$ inequality \cite{brunner2008partial}, which can be written as
\begin{equation}
    I^4_{4422} \leq 0,
\end{equation}
with
\begin{equation}
\begin{split}
    I^4_{4422} &= I^{(1,2;1,2)}_{\text{CH}} + I^{(3,4;3,4)}_{\text{CH}} - I^{(2,1;4,3)}_{\text{CH}} - I^{(4,3;2,1)}_{\text{CH}} \\
                & - p_{A}(1|2) - p_{A}(1|4) - p_{B}(1|2) - p_{B}(1|4),
\end{split}
\end{equation}
where $I^{(i,j;u,v)}_{\text{CH}} = p(1,1|i,u) + p(1,1|j,u) + p(1,1|i,v) - p(1,1|j,v) - p_{A}(1|i) - p_{B}(1|u)$.

The reason for choosing this Bell inequality is that the threshold detection efficiency for attaining the loophole-free regime for \(I^4_{4422}\) is \(0.7698\) and \(0.618\), for maximally and partially entangled states, respectively \cite{vertesi2010closing}. That is, the \(I^4_{4422}\) inequality requires lower threshold detection efficiencies than the CHSH inequality, for which these thresholds are \( 0.828\) and \( 2/3\) \cite{eberhard1993background}, respectively. Therefore, the hope is that this advantage can be translated into a similar advantage in the threshold detection efficiency for DI-QKD.

We used the quantum realization $Q_{4422}$ given in the supplementary material of Ref.~\cite{vertesi2010closing}, consisting of the maximally entangled two-ququart state
\begin{equation}
\label{mes}
    \ket{\psi} = \frac{1}{2}\left(\ket{11}+\ket{22}+\ket{33} +\ket{44}\right),
\end{equation}
and four measurements with two outcomes. Each measurement is a projector defined by four coefficients $c_i$ as follows:
\begin{equation}\label{eq.normalization}
    \sum^4_{i,j=1} c_i c^*_j \ket{i} \bra{j},
\end{equation}
with $\sum^4_{i=1} |c_i|^2 = 1$. The coefficients for Alice's measurements $x \in \{1,2,3,4\}$ are
\begin{equation}
\begin{split}
    1 &= (-0.2816,-0.2816,0.9159,0.0499),\\
    2 &= (-0.5438,0.5438,0.5625,-0.3035),\\
    3 &= (0.2816,0.2816,0.9159,0.0499),\\
    4 &= (0.5438,-0.5438,0.5625,-0.3035),\\
\end{split}
\end{equation}
and for Bob's measurements $y \in \{1,2,3,4\}$ are
\begin{equation}
\begin{split}
    1 &= (-0.2816,0.2816,0.9159,-0.0499),\\
    2 &= (-0.5438,-0.5438,0.5625,0.3035),\\
    3 &= (0.2816,-0.2816,0.9159,-0.0499),\\
    4 &= (0.5438,0.5438,0.5625,0.3035).\\
\end{split}
\end{equation}
Since each measurement is a projector, it has only two eigenvalues $\lambda \in \{0,1\}$. For the QKD protocol, we consider that the output `1' corresponds to the eigenvalue $\lambda = 1$, while the output `2' corresponds to $\lambda = 0$.
$Q_{4422}$ does not reach the maximum quantum violation of the \(I^4_{4422}\) inequality, which is achieved with pairs of real qubits and degenerate measurements \cite{pal2009}.

\begin{figure}[t]
    \centering
    \includegraphics[scale=0.6]{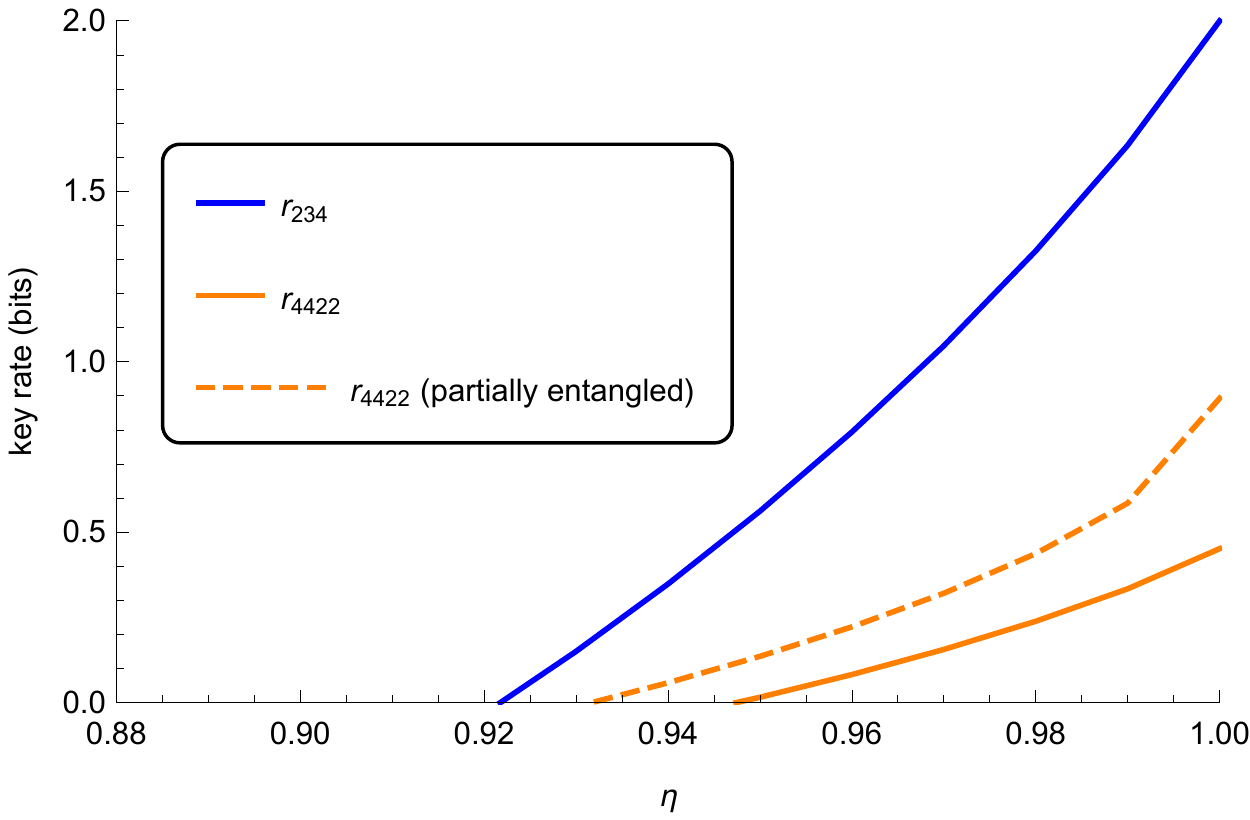}
    \caption{Secret key rates versus detection efficiency \(\eta\). Comparison between the secret key rate obtained with \(Q_{234}\) (blue line), \(Q_{4422}\) (solid orange line), and the realization described in Section \ref{sec.B} (dashed orange line). The corresponding thresholds to distill a secret key are 0.9218, 0.9474 and 0.9317, respectively.}
    \label{fig:rvsEta}
\end{figure}

\begin{figure}[tbh]
    \centering
    \includegraphics[scale=0.6]{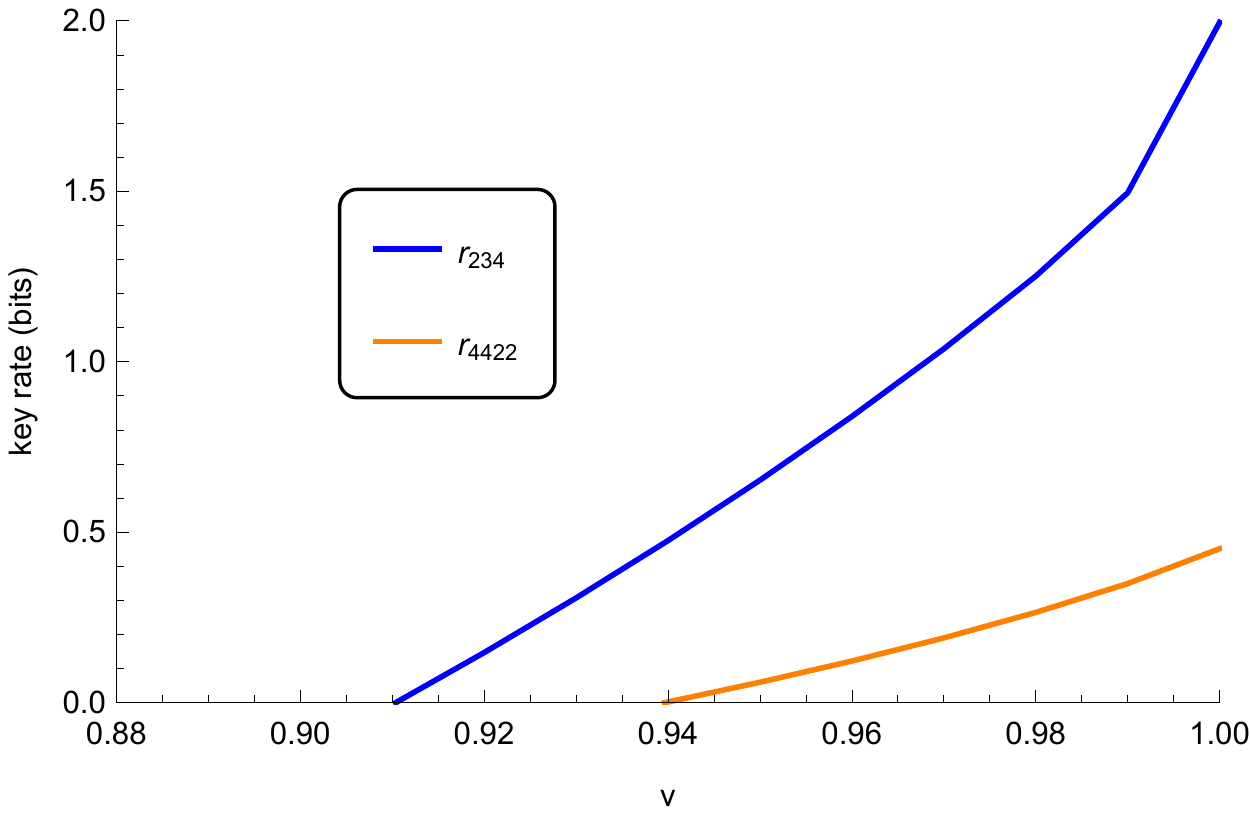}
    \caption{Secret key rates versus visibility \(V\). Comparison between the secret key rate obtained with the \(Q_{234}\) realization (blue) and the \(Q_{4422}\) one (orange). The corresponding thresholds to distill a secret key are 0.9104 and 0.9396, respectively.}
    \label{fig:rvsV}
\end{figure}

The key rate of the resulting QKD protocol, as a function of the detection efficiency, is presented in  Fig.~\ref{fig:rvsEta} (solid orange line), while Fig.~\ref{fig:rvsV} (orange line) shows the key rate as a function of the visibility. 

The thresholds to distill a secret key are \(\eta \geq 0.9474\) and \(V \geq 0.9396\). Note that $Q_{4422}$ does not reach the value of $1$~bit for the key rate, even with $\eta =1$ and $V = 1$. Instead, the maximum key rate obtained is $0.4516$~bits.


\subsection{Protocol based on $I^4_{4422}$ using partially entangled states}\label{sec.B}


We also investigated the performance of the protocol using partially entangled states. For that, we searched for the quantum realization that maximizes the value \(I^4_{4422}\) for a particular class of ququart-ququart entangled states and measurements, optimized for each $\eta$. 
Specifically, we use the parametrization given in Ref.~\cite{vertesi2010closing}. This parametrization is
\begin{equation}
    \ket{\psi(\epsilon)} = \sqrt{\frac{1-\epsilon^2}{3}}\left(\ket{11}+\ket{22}+\ket{33}\right) +\epsilon\ket{44}
\end{equation}
and gives a maximally entangled ququart-ququart state for $\epsilon = 1/2$. As in the previous case, the measurements performed are projectors. Alice's measurements $x \in \{1,2,3,4\}$ have the coefficients 
\begin{equation}
\begin{split}
    1 &= (-u, -u, \Vec{p}_1),\\
    2 &= (-v, v, \Vec{p}_2),\\
    3 &= (u, u, \Vec{p}_1),\\
    4 &= (v, -v, \Vec{p}_2),
\end{split}
\end{equation}
and for Bob's measurements $y \in \{1,2,3,4\}$ are
\begin{equation}
\begin{split}
    1 &= (-u, u, \Vec{q}_1),\\
    2 &= (-v, -v, \Vec{q}_2),\\
    3 &= (u, -u, \Vec{q}_1),\\
    4 &= (v, v, \Vec{q}_2),
\end{split}
\end{equation}

where $\Vec{p}_i = (p_{i1},p_{i2})$ and $\Vec{q}_i = (q_{i1},q_{i2})$, with $i \in \{1,2\}$. The normalization condition $\sum^4_{i=1} |c_i|^2 = 1$ is introduced as a constrain in the optimization.

\begin{figure}[tbh]
    \centering
    \includegraphics[scale=0.6]{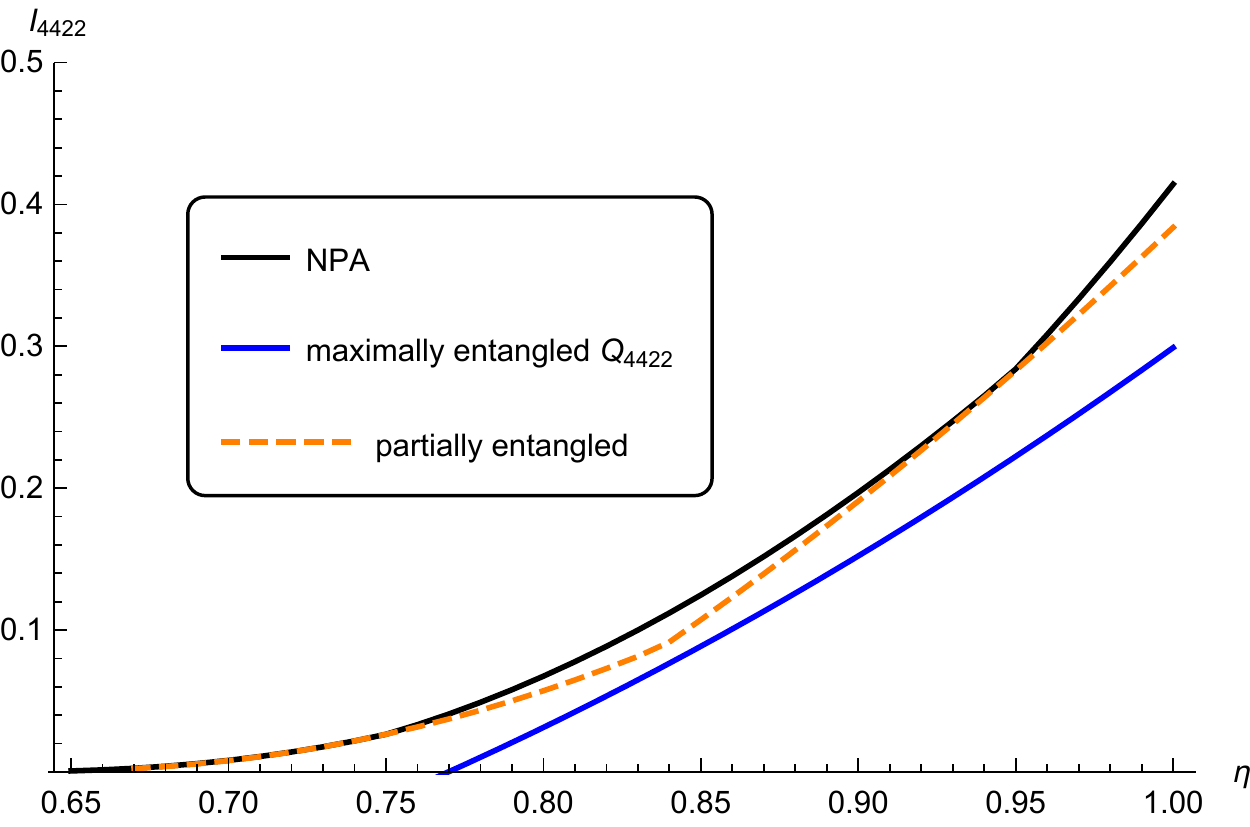}
    \caption{Comparison between the violations of the \(I_{4422}\) inequality with maximally (blue) and partially entangled states (orange). In addition, the upper bound was calculated using the NPA hierarchy (black). The parametrization used is close to be optimal for \(\eta \leq 0.75\) and \(0.90 \leq \eta \leq 0.95\). For any detection efficiency $\eta$ the partially entangled states outperformed the maximally entangled ones.}
    \label{fig:I4422vsEta}
\end{figure}

We found that stronger violations than the ones obtained with maximally entangled states can be reached using partially entangled states. This is shown in Fig. \ref{fig:I4422vsEta}, where the violations obtained are compared with the bounds to the maximal quantum violation of the $I^4_{4422}$ inequality computed using the Navascu\'es-Pironio-Ac\'{\i}n (NPA) hierarchy \cite{navascues2008convergent}.

Therefore, better bounds for \(H(A|E)\) were achieved. This allowed us to improve the threshold to distill a secret key in this scenario. The threshold that we obtained using partially entangled states was $\eta = 0.9317$. To accomplish this improvement, we had to also optimize Bob's measurement $B_{n+1}$, targeting the minimum \(H(A_1|B_{n+1})\). 


\subsection{Protocol based on the (2,3,4) Bell inequality using maximal ququart-ququart entanglement}


The second protocol is constructed around the following Bell inequality with three inputs and four outputs: 
\begin{equation}
  I_{234} \leq 8,  
\end{equation}
with
\begin{widetext}
\begin{equation}\label{eq.c1}
    \begin{split}
        I_{234} &= p(1,1|1,1) + p(1,2|1,1) + p(2,1|1,1) + p(2,2|1,1) + p(3,3|1,1) + p(3,4|1,1) + p(4,3|1,1) + p(4,4|1,1)  \\
            &+p(1,1|1,2) + p(1,2|1,2) + p(2,3|1,2) + p(2,4|1,2) + p(3,1|1,2) + p(3,2|1,2) + p(4,3|1,2) + p(4,4|1,2)  \\
            &+p(1,1|2,1) + p(1,3|2,1) + p(2,1|2,1) + p(2,3|2,1) + p(3,2|2,1) + p(3,4|2,1) + p(4,2|2,1) + p(4,4|2,1)  \\
            &+p(1,1|2,2) + p(1,3|2,2) + p(2,2|2,2) + p(2,4|2,2) + p(3,1|2,2) + p(3,3|2,2) + p(4,2|2,2) + p(4,4|2,2)  \\
            &+p(1,1|1,3) + p(1,2|1,3) + p(2,3|1,3) + p(2,4|1,3) + p(3,3|1,3) + p(3,4|1,3) + p(4,1|1,3) + p(4,2|1,3)  \\
            &+p(1,1|2,3) + p(1,3|2,3) + p(2,2|2,3) + p(2,4|2,3) + p(3,2|2,3) + p(3,4|2,3) + p(4,1|2,3) + p(4,3|2,3)  \\
            &+p(1,1|3,1) + p(1,4|3,1) + p(2,1|3,1) + p(2,4|3,1) + p(3,2|3,1) + p(3,3|3,1) + p(4,2|3,1) + p(4,3|3,1)  \\
            &+p(1,1|3,2) + p(1,4|3,2) + p(2,2|3,2) + p(2,3|3,2) + p(3,1|3,2) + p(3,4|3,2) + p(4,2|3,2) + p(4,3|3,2)  \\
            &+p(1,2|3,3) + p(1,3|3,3) + p(2,1|3,3) + p(2,4|3,3) + p(3,1|3,3) + p(3,4|3,3) + p(4,2|3,3) + p(4,3|3,3). 
    \end{split}
\end{equation}
\end{widetext}
This inequality was introduced in \cite{cabello2001all} and is tight in the two-party, three-setting, four-measurement or
(2,3,4) scenario \cite{gisin2007pseudo}. Its maximum quantum violation requires pairs of ququarts. 

The reason for choosing this inequality is that it is the simplest bipartite Bell inequality in which the maximum quantum violation equals the nonsignaling bound \cite{aolita2012}. 

We chose a quantum realization that leads to the maximum violation. This realization $Q_{234}$ is defined as follows. 
Each party has  a ququart. The initial state of the pair is the maximally entangled state given in Eq.~(\ref{mes}).
Each ququart can be seen as a pair of qubits. 
Then, each local measurement can be seen as a pair of compatible measurements on the corresponding qubit-qubit local system. 
Specifically, if we relabel the basis of Alice and Bob as
\begin{equation}
    \begin{split}
        \ket{1} & \xrightarrow{} \ket{0}  \ket{0}, \\
        \ket{2} & \xrightarrow{} \ket{0}  \ket{1}, \\
        \ket{3} & \xrightarrow{} \ket{1}  \ket{0}, \\
        \ket{4} & \xrightarrow{} \ket{1}  \ket{1}, \\
    \end{split}
\end{equation}
then Alice's measurements can be written as
\begin{equation}\label{eq.6}
    \begin{split}
        1 &= (\sigma_z^{(b)},\sigma_x^{(a)}),\\
        2 &= (\sigma_z^{(a)},\sigma_x^{(b)}),\\
        3 &= (\sigma_z^{(a)} \otimes \sigma_z^{(b)},\sigma_x^{(a)} \otimes \sigma_x^{(b)}).
    \end{split}
\end{equation}
Similarly, Bob's measurements can be written as
\begin{equation}\label{eq.7}
    \begin{split}
        1 &= (\sigma_z^{(d)},\sigma_z^{(c)}),\\
        2 &= (\sigma_x^{(c)},\sigma_x^{(d)}),\\
        3 &= (\sigma_x^{(c)} \otimes \sigma_z^{(d)},\sigma_z^{(c)} \otimes \sigma_x^{(d)}),
    \end{split}
\end{equation}
where \(\sigma_i^{k}\) denotes the \(i=\{x,y,z\}\) Pauli matrix for qubit \(k = \{a,b,c,d\}\). The qubits $\{a,b\}$ correspond to Alice, while qubits $\{c,d\}$ to Bob.

The local observables correspond to the three rows (Alice's observables) and columns (Bob's) of the Peres-Mermin table \cite{peres1990incompatible,mermin1990simple}.
Previous DI-QKD protocols have used this realization \cite{horodecki2010contextuality,jain2020parallel}.


In addition, the pair of outcomes $(\pm 1,\pm 1)$ that each measurement yields can also be mapped as $(+1,+1) \xrightarrow{} 1$, $(+1,-1) \xrightarrow{} 2$, $(-1,+1) \xrightarrow{} 3$ and $(-1,-1) \xrightarrow{} 4$. 

The key rates of this protocol are presented in Figs. \ref{fig:rvsEta} and \ref{fig:rvsV}. We obtained the following thresholds to distill a secret key, \(\eta \geq 0.9218\) and \(v \geq 0.9104\). Our detection efficiency threshold is lower than the one obtained with the CHSH-based protocol using a maximally entangled state (0.924). However, it is still beyond the scope of the current technology. On the other hand, our visibility threshold, under the assumption of perfect detection efficiency, shows that the experiment reported in Ref.~\cite{yang2005all} has good enough visibility to distill a secret key (\(v \approx 0.95\)). 


\subsection{Protocol based on the (2,3,4) Bell inequality using qutrit-qutrit partially entangled states}\label{sec.d}


We optimized the inequality \(I_{234}\) over the family of states given by
\begin{equation}
\begin{split}
    \ket{\psi} = &\cos{\theta_1} \cos{\theta_2} \ket{11} +\cos{\theta_1} \sin{\theta_2} \ket{22} \\
    &+\sin{\theta_1} \cos{\theta_2} \ket{33}  +\sin{\theta_1} \sin{\theta_2} \ket{44}.
\end{split}
\end{equation}
In the most general case, each four-outcome measurement is specified by 15~parameters. However, using 15~variables per measurement would make the parameter space extremely large and impractical to optimize. Therefore, we opted for trying an ansatz that involved less parameters. Concretely, we tried three different parametrizations for the measurements. Our first parametrization was given by projections in the plane for each qubit, 
\begin{equation}
    \begin{split}
        i &= (\cos{\alpha}~\sigma_z + \sin{\alpha}~\sigma_x)_a \otimes (\cos{\beta}~\sigma_z + \sin{\beta}~\sigma_x)_b ,\\
        j &= (\cos{\gamma}~\sigma_z + \sin{\gamma}~\sigma_x)_c \otimes (\cos{\delta}~\sigma_z + \sin{\delta}~\sigma_x)_d,
    \end{split}
\end{equation}
where $i=1,2,3$ are Alice's measurements and $j=1,2,3$ are Bob's.

The second parametrization was the one introduced in Refs.~\cite{collins2002bell,acin2002quantum}. For this parametrization is convenient to define the projectors into each subspace for Alice and Bob. Specifically, we denote by $x$ the measurements that Alice perform and with $y$ the measurements of Bob. Moreover, the outputs of the measurements are $a$ for Alice and $b$ for Bob. The projectors are then given by
\begin{equation}
    \begin{split}
        \Pi_x^a &= V(\Vec{\phi_x})^{\dagger} \ket{a} \bra{a} V(\Vec{\phi_x}) ,\\
        \Pi_y^b &= V(\Vec{\varphi_y})^{\dagger} \ket{b} \bra{b} V(\Vec{\varphi_y}),
    \end{split}
\end{equation}
where \(V(\Vec{\phi_x}) = U_{\text{FT}}~U(\Vec{\phi_x})\) and \(V(\Vec{\varphi_y}) = U^*_{\text{FT}}~U(\Vec{\varphi_y})\). The unitary operators $U(\Vec{\phi_x})$ and  $U(\Vec{\varphi_y})$ have nonzero elements only in their diagonal. The elements in the diagonal are \(\exp{i \phi_x(k)}\) for Alice and \(\exp{i \varphi_y(l)}\) for Bob. Also, $\phi_x(k)$ and $\varphi_y(l)$ refer to the components of the real vectors $\Vec{\phi}_x$ and $\Vec{\varphi}_y$, respectively. \(U_{\text{FT}}\) is the Fourier transform. Under this prescription, the probability is
\begin{equation}
    p(a,b|x,y) = \tr{\Pi_x^a \otimes \Pi_y^b \rho}.
\end{equation}

The third parametrization uses pairs of compatible measurements as in Eqs.~(\ref{eq.6}) and (\ref{eq.7}). However, now each operator, aside from the identity, is replaced by \(\cos{\alpha}~\sigma_z + \sin{\alpha}~\sigma_x\).

The first and second parametrizations do not reach any violations of the \(I_{234}\) inequality. The third parametrization achieved values of \(I_{234} \geq 8\). Howbeit, these are smaller than the violations obtained by maximally entangled states together with measurements in Eqs.~(\ref{eq.6}) and (\ref{eq.7}). As a consequence, none of the models is able to improve the key rate presented in Fig.~\ref{fig:rvsEta}.

\begin{figure}[t]
    \centering
    \includegraphics[scale=0.6]{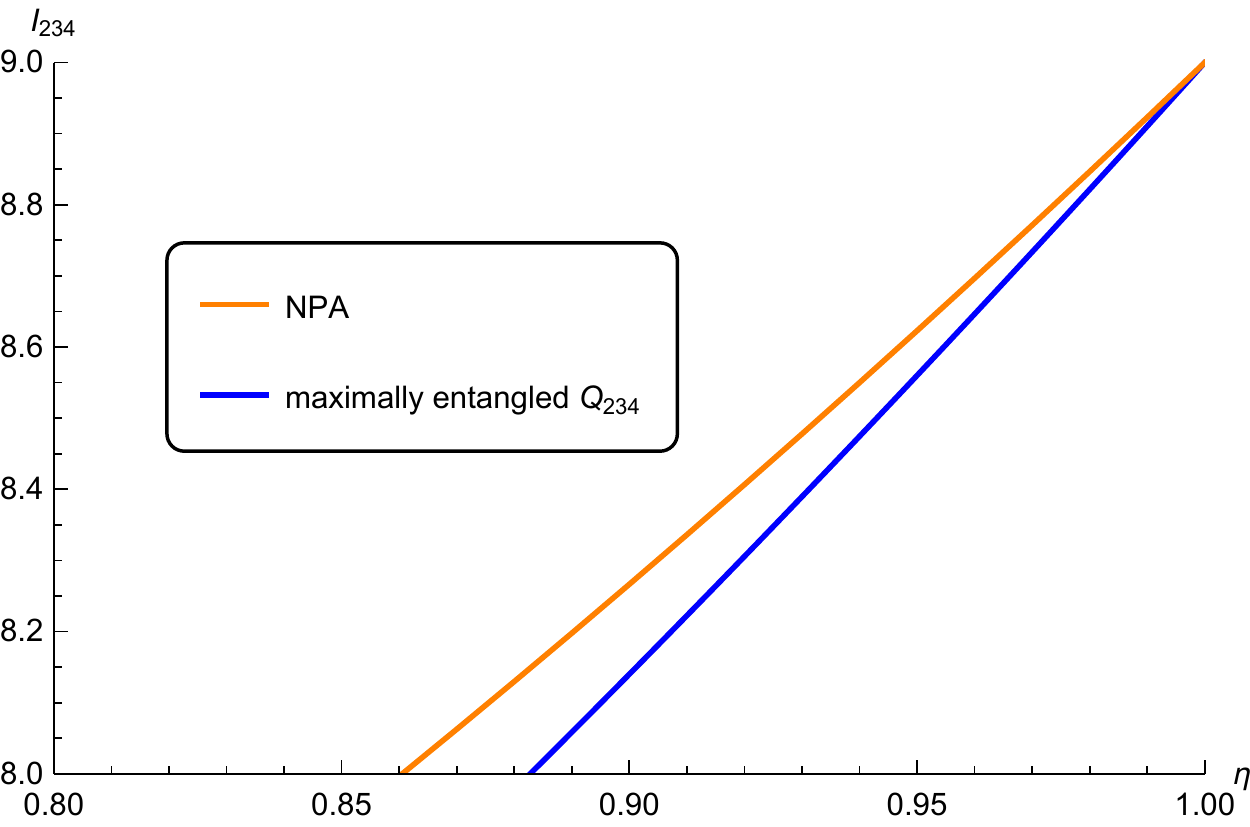}
    \caption{Comparison of the violation of the \(I_{234}\) inequality with maximally entangled states (blue) and the maximum allowed by quantum mechanics (orange). The gap between both realizations increases when detectors are less efficient.}
    \label{fig:I234vsEta}
\end{figure}

We speculate that for $\eta < 1$, there are quantum realizations that are better than the one we used $Q_{234}$. Our guess is supported by the fact that the NPA hierarchy predicts stronger violations of the $I_{234}$ when $\eta < 1$. These quantum realizations can potentially improve the detection efficiency threshold that we presented in Fig. \ref{fig:rvsEta}. Further research is required to find a suitable parametrization and identify the state and measurements that reach the violations of $I_{234}$ predicted by the NPA hierarchy (Fig.~\ref{fig:I234vsEta}).


\section{Conclusions}


We have presented two DI-QKD protocols whose security relies on Bell inequalities with more than two inputs and two outputs. Both protocols were based on behaviors that display features that are potentially interesting for DI-QKD.
Our best results were obtained with the protocol based on a behavior using maximally entangled states violating the (2,3,4) Bell inequality. We found that this protocol requires a lower detection efficiency than the protocol based on maximally entangled states  violating the CHSH inequality. 
Regarding visibility, we have noticed that previous experiments based on $Q_{234}$ could be used to distill a cryptographic key (if $\eta$ would have been $1$). 

Our results could be improved using more general partially entangled states and noise pre-procesing \cite{woodhead2020device,ho2020noisy} (see Section \ref{sec.d}). We will investigate these possibilities in a future work.

Summing up, we have addressed the problem of the detection efficiency in DI-QKD protocols from a different angle than previous works \cite{woodhead2020device,sekatski2020device,schwonnek2020robust,ho2020noisy}. Although, our findings are still not capable to close the gap between theoretical and practical security, they show that DI-QKD protocols indeed benefit from the usage of Bell inequalities with more than two inputs and two outputs and should stimulate further research in this direction.


\begin{acknowledgments}
The authors thank P.\ Brown and D.\ Garc\'{\i}a for enlightening discussions.  A.C.\ is supported by Project Qdisc (Project No.\ US-15097), with FEDER funds, MINECO Project No.\ FIS2017-89609-P, with FEDER funds, by MINECO (Project No.\ PCI2019-111885-2). A.P.\ would like to acknowledge Swedish Research Council. J.G.U. is supported by project HYPER-U-P-S.  Project  HYPER-U-P-S  has  received  funding  from  the QuantERA ERA-NET  Cofund  in  Quantum  Technologies  implemented  within  the  European  Union’s  Horizon  2020  Programme. 
\end{acknowledgments}


\appendix

\section{Methods} \label{app.1}

Here, we summarize the  method to bound the quantum conditional entropy \(H(A|E)\) of Brown {\em et al.} \cite{brown2020computing}. They introduced a family of iterated mean quantum R\'{e}nyi divergences that is applicable on device-independent tasks. Using this technique, it is possible to lower bound device-independent random number generation rates, as well as secret key rates of DI-QKD protocols. The required quantum R\'{e}nyi divergence is obtained after the following optimization (Lemma~1 in Ref.~\cite{brown2020computing}):

\textbf{Lemma} \textit{ Let \(\ket{\psi}\bra{\psi} \in D(Q_AE)\), \(\{M_a\}_{a\in A}\) be a POVM on \(Q_A\) and \(\rho_{AE} = \Sigma_a \ket{a}\bra{a}\otimes\rho_E(a)\) be a cq-state where \(\rho_E(a)= \tr_{Q_A}[(M_a \otimes \mathds{1}) \ket{\psi}\bra{\psi}]\). Then, for each \(k \in \mathds{N}\), we have}
\begin{equation}
    H^{\uparrow}_{(\alpha_k)} = \frac{\alpha_k}{1-\alpha_k} \log Q^{DI}_{(\alpha_k)},
\end{equation}
\textit{where}
\begin{widetext}
\begin{equation}\label{eq.1}
    \begin{split}
        Q^{DI}_{\alpha_k} = \underset{V_{j,a}:1\leq j \leq k ,a\in A}{\text{max}} ~ &\sum_{a} \tr{(M_a \otimes \frac{V_{1,a}+V^*_{1,a}}{2}) \ket{\psi}\bra{\psi}}  \\
        \text{s.t.} &\sum_{a} V^*_{k,a} V_{k,a} \leq I_E\\
            & V_{1,a}+V^*_{1,a} \geq 0 ~~~~~~~~~~~~~~~~~\textit{for all a \(\in\) A}\\
            & 2V^*_{i,a}V_{i,a} \leq V_{i+1,a}+V^*_{i+1,a} ~~\textit{for all \(1\leq i \leq k-1\) and a \(\in\) A}
    \end{split}
\end{equation}
\end{widetext}

The family of iterated mean quantum R\'{e}nyi divergences is indexed by the parameter \(\alpha_k = 1+ \frac{1}{2^k-1}\), with \(k\) being a positive integer. The optimization in Eq.~(\ref{eq.1}) is not yet yielding a device-independent entropy, as it is performed in a particular Hilbert space. In order to compute the entropy \(H^{\uparrow}_{(\alpha_k)}\) in a device-independent way, the optimization can be relaxed to a semi-defined program through the NPA hierarchy \cite{navascues2008convergent}. Whether or not the relaxation of \(H^{\uparrow}_{(\alpha_{K} )}\) converges to \(H(A|E)\) for \(k \xrightarrow{} \infty\), it is still an open question. This technique does not rely on any Bell inequality to estimate \(H^{\uparrow}_{(\alpha_k)}\); instead, it uses the complete behavior of the system as a constrain of Eq.~(\ref{eq.1}), in contrast to the CHSH-based protocol \cite{pironio2009device}.

Since $H^{\uparrow}_{(\alpha_k )}$ constitutes a valid lower bound of $H(A|E)$ for any value of $k$, we can choose the $k$ that fits better our hardware. The calculations of $r_{4422}$ were performed with $k=2$ up to the second level of the NPA hierarchy. The certificates required approximately 30 GB of RAM. The calculations for $r_{234}$ proved to be considerable more demanding given the high number of inputs and outputs. At first instance, we performed the calculations with $k=2$, but we ran out of memory to include the necessary monomials to ensure $H^{\uparrow}_{(\alpha_{k+1} )} \geq H^{\uparrow}_{(\alpha_k )}$. Thus, this first estimation was outperformed by the result with $k=1$. Finally, the results presented for $r_{234}$ were all calculated with $k=1$ (applying the dilation theorem in the supplementary information of Ref.~\cite{brown2020computing}), the NPA hierarchy at level~2, and adding as many monomials as our hardware allowed us. The certificates required around 180 GB of RAM. All the SDPs in this work were constructed with the Python package `ncpol2sdpa' and solved with Mosek \cite{wittek2015algorithm,mosek}.

\bibliographystyle{apsrev4-2}


%

\end{document}